%% file: main.tex
\documentclass{article}
\usepackage[utf8]{inputenc}
\usepackage{authblk}
\usepackage{setspace}
\usepackage[margin=1.25in]{geometry}
\usepackage{graphicx}
\usepackage{diagbox}
\graphicspath{ {./figures/} }
\usepackage{subcaption}
\usepackage{amsmath}
\usepackage{amsfonts}

\usepackage[style=ieee, 
citestyle=numeric-comp,
sorting=none]{biblatex}
\title{Breast Cancer Induced Bone Osteolysis Prediction Using Temporal Variational Auto-Encoders}

\author[1$\dag$*]{Wei Xiong}
\author[1$\dag$]{Neil Yeung}
\author[2]{Shubo Wang}
\author[3]{Haofu Liao}
\author[2]{Liyun Wang}
\author[1]{Jiebo Luo}

\affil[1]{Department of Computer Science, University of Rochester, Rochester, USA.}
\affil[2]{Department of Mechanical Engineering, University of Delaware, USA.}
\affil[3]{Amazon Web Services, USA.}
\affil[$\dag$]{These authors contributed equally to this work.}
\affil[*]{Corresponding author. Email: wei.xiong@rochester.edu}

\date{}

\begin{document}

\maketitle

\input{abstract}

\input{introduction}

\input{results}

\input{discussion}

\input{method}

\input{animalstudies}

\input{acknowledgements}

\printbibliography

\end{document}

%% file: abstract.tex
\begin{abstract}

\noindent\textit{Objective and Impact Statement}. We adopt a deep learning model for bone osteolysis prediction on computed tomography (CT) images of murine breast cancer bone metastases. Given the bone CT scans at previous time steps, the model incorporates the bone-cancer interactions learned from the sequential images and generates future CT images. Its ability of predicting the development of bone lesions in cancer-invading bones can assist in assessing the risk of impending fractures and choosing proper treatments in breast cancer bone metastasis.
\textit{Introduction}. Breast cancer often metastasizes to bone, causes osteolytic lesions, and results in skeletal related events (SREs) including severe pain and even fatal fractures. Although current imaging techniques can detect macroscopic bone lesions, predicting the occurrence and progression of bone lesions remains a challenge.
\textit{Methods}. We adopt a temporal variational auto-encoder (T-VAE) model that utilizes a combination of variational auto-encoders and long short-term memory networks to predict bone lesion emergence on our micro-CT dataset containing sequential images of murine tibiae. Given the CT scans of murine tibiae at early weeks, our model can learn the distribution of their future states from data. 
\textit{Results}. We test our model against other deep learning-based prediction models on the bone lesion progression prediction task. Our model produces much more accurate predictions than existing models under various evaluation metrics.
\textit{Conclusion}. We develop a deep learning framework that can accurately predict and visualize the progression of osteolytic bone lesions. It will assist in planning and evaluating treatment strategies to prevent SREs in breast cancer patients. 

\end{abstract}

%% file: introduction.tex
\section{Introduction}
Bone is among the most common sites of cancer metastasis, whereby primary cancers originating from other places such as breast, prostate, colon, and kidney spread to various bones including the spine, hip and skull. Osteolytic bone lesions, which result from pathological bone loss due to tumor invasion, are developed in around 75\%\ patients with stage IV breast cancer, the most common non-skin cancer among women in the United States \cite{cancer_stat}. The destruction of bone is driven by the “vicious” cycle between breast cancer cells and bone reabsorbing osteoclasts, in which one would reinforce the activity of the other \cite{metastasis,bone_events}. As the result, cancer patients with bone metastasis can suffer from more than two adverse skeletal related events (SREs) such as severe bone pain, spinal cord compression, and bone fractures during the course of the disease. It is difficult to treat SREs \cite{bone_events}. 

In current practice, computed tomography (CT) is routinely employed to evaluate bone structure, which shows 95\%\ specificity in detecting bone metastasis with relatively low sensitivity (73\%)~\cite{imaging_dia}. To improve the accuracy, CT has been combined with other imaging modalities such as positron-emission tomography (PET) or magnetic resonance imaging (MRI) to detect cancer bone metastasis \cite{imaging_dia}. Despite the advance in detecting cancer bone metastasis using these hybrid imaging approaches, it is still challenging to predict fracture risk, which is crucial for planning treatments and improving clinical outcomes. While prophylactic stabilization has been suggested to increase overall and immediate postoperative survival, the well-accepted Mirels' scoring scheme used in clinical practice lacks reproducibility and specificity (13\%-50\%) for long bones with metastatic lesions\cite{damron2020fracture}. Several CT based fracture risk assessment studies  have attempted to address this challenge, for example, by conducting rigidity analysis using 2D trans-axial CT images~\cite{trans-axial-rigidity} or finite element analysis using 3D volumetric images~\cite{FEM2012, FEM2018, eggermont2018can}. These approaches, tested on CT images at single time points, could potentially be more useful if future bone structure were accurately predicted. Prior studies have also investigated prediction of SREs using machine learning methods but limited efforts on early assessment of bone fracture in metastatic bone disease have been made.

Although sequential clinical images may be captured for breast cancer patients, there have been few attempts to utilize prior scans in assessing bone lesions and the associated fracture risk. In a preclinical study, the accuracy of bone metastasis prediction based on one prior dual-modality image was found to be approximately 85\%\ in a rat breast cancer metastasis model \cite{pet_ct_mri_prediction}. Since bone metastasis lesions often progress with time, a temporal sequence of images may offer more cues leading to better detection of bone lesions. To this end, we have performed a preclinical imaging study in mice to test if a sequence of CT images could be used to predict breast cancer induced bone destruction utilizing deep learning. The intra-tibial cancer inoculation model showed the full progression of bone destruction, from trabecular bone loss within the bone marrow to full-thickness perforation of bone cortex, similar to clinical observations in breast cancer patients \cite{murinemodel}. 

Based on the preclinical study and the recent progress on medical image analysis with deep learning methods \cite{ronneberger2015u,GP-GAN,li2019patch,li2020structured}, we showed that deep learning methods can be an effective approach to predicting bone metastasis. To achieve this, in this study, we collected a dataset composed of micro-computed tomography (microCT) scans of murine skeleton, which were taken at 3-6 successive time points while breast cancer metastasis and bone lesions progressed.
Specifically, we aimed to predict and visualize the progression process of bone lesions using a deep learning framework. We benchmarked our model against  state-of-the-art future prediction deep models, the 3D-Generative Adversarial Networks (3D-GAN) \cite{GP-GAN, xiong2018learning}, the Convolutional Long Short Term Memory Network (C-LSTM) \cite{zhang2019spatiotemporal}, and the PredRNN \cite{wang2021predrnn}. 

The main novelty of our approach is that we modeled the bone lesion prediction problem as a video prediction problem \cite{svg-lp, xiong2018learning}. Inspired by the success of recent natural video prediction methods \cite{svg-lp}, we adopted a temporal variational auto-encoder (T-VAE) \cite{kingma2013auto, vae} model that captures both the spatial and temporal patterns of cancer-induced bone osteolysis. We used variational auto-encoders to model the distribution of the future states. Moreover, we proposed an edge-aware loss that encourages our model to pay more attention to the valid pixels in the sparse CT slices, as an effective way to address the challenging class imbalance between the foreground class pixels (e.g. the bone) and the background class pixels within the CT image data. Given the bone CT images taken at the first three weeks, our model was shown to predict the progression of bone lesions at the fourth week with an accuracy significantly higher than the benchmark models.  

%% file: results.tex

\begin{figure}[t]
    \centering
    \includegraphics[width=0.75\columnwidth]{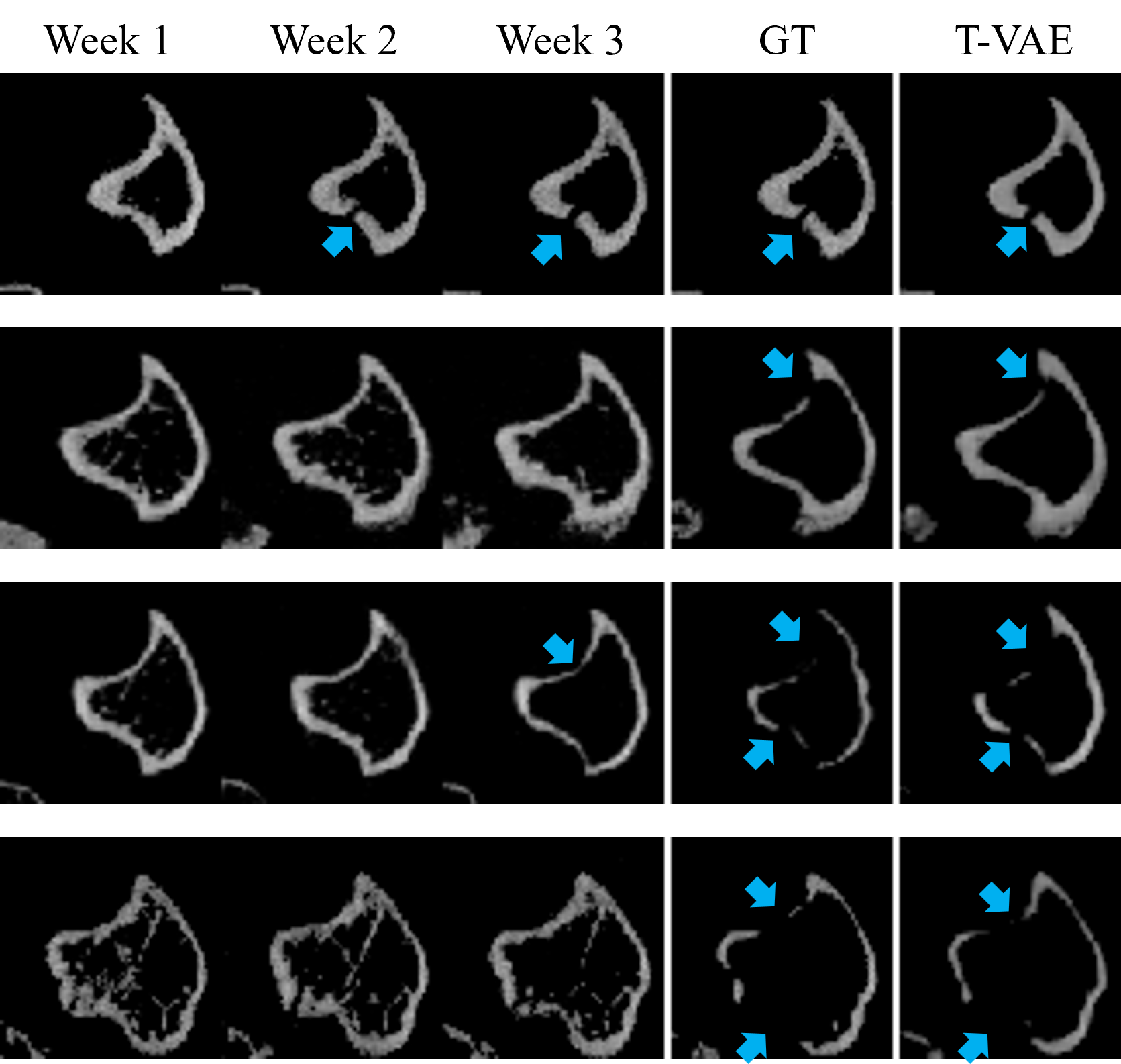}
    \caption{Each row displays the first three weeks of bone CT slices (input), the ground truth image in week 4 and \emph{our} predicted result in week 4. Blue arrows indicate osteolytic lesions.}
    \label{demo}
\end{figure}

\section{Results}

\subsection{Task Definition} The specific task in this work is that given the first three weeks of murine bone CT scans, we predict and generate CT scan images at the fourth week. Figure~\ref{demo} shows the visualization of the input transverse/axial CT slices, the ground-truth CT slice at the fourth week and the generated CT slice by our Temporal VAE (T-VAE) model. Comparing our generated images with the ground-truth data,  the visual results demonstrate that our model can predict the bone lesions accurately.

\subsection{Evaluation Metrics}
We evaluate the quality of the generated predictions utilizing the quantitative metrics of peak signal to noise ratio (PSNR), structural similarity index measurement (SSIM), and Learned Perceptual Image Patch Similarity (LPIPS) \cite{zhang2018unreasonable} score of the generated predicted fourth frame against the ground truth fourth frame.

\begin{figure*}[t]
    \centering
    \includegraphics[width=\textwidth]{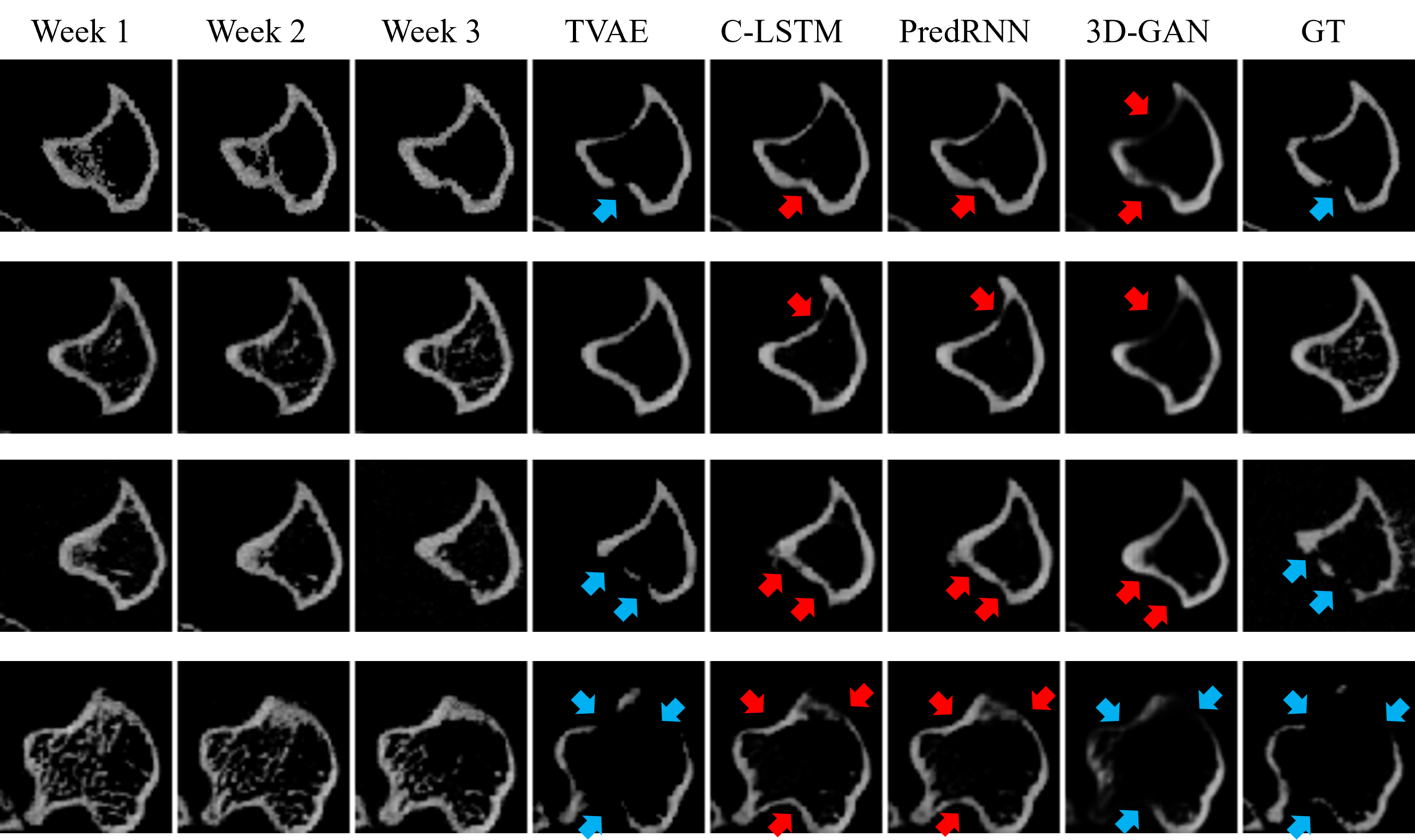}
    \caption{Each row displays the first three weeks of bone CT slices (input), followed with the \emph{predicted} results from T-VAE (2D), C-LSTM, PredRNN, 3D-GAN, and ground truth image in week 4. Blue arrows indicate lesions in the ground-truth image or correct osteolytic lesion predictions. Red arrows indicate wrong lesion predictions. Please pay special attention to the boundary regions of the bone (i.e., the tibial cortical bone), which are the primary regions to determine whether their is lesion or not.  Qualitatively, \emph{our} model T-VAE provides the best results among all the models.}
    \label{comparison}
\end{figure*}

\subsection{Comparison with Other Models}
We compare our T-VAE model with several state-of-the art future frame prediction models: the 3D Generative Adverserial Networks (3D-GAN) \cite{xiong2018learning}, the Convolutional LSTM (C-LSTM) \cite{convlstm} and the PredRNN \cite{wang2021predrnn} for the slice prediction task. 3D-GAN learns a distribution of future frames with a pair of generator and discriminator composed of 3D-convolutions \cite{goodfellow2014generative, dcgan, xiong2020fine}. C-LSTM first projects images into feature maps with convolution operations, then learns the temporal information with an LSTM model. PredRNN uses spatiotemporal memory flow to build the architecture for accurate future prediction.

\input{tables/compare}

\textbf{Qualitative Comparison.} Figure \ref{comparison} shows the visual results generated by different prediction models.  In Figure \ref{comparison}, there is no lesion in the slices of the first three weeks but lesion may occur in the fourth week. This makes the prediction very challenging. 
On such cases, all the compared models fail to make accurate bone lesion predictions. In contrast, our model is able to capture subtle temporal progression patterns in the first three weeks and predict the state of the bone image in the fourth week accurately. Notably, 3D-GAN makes a few correct lesion predictions. However, it tends to over-predict the lesion, i.e., the lesion region is much larger than the ground-truth lesion or lesion is predicted but there is no lesion in the ground-truth.

\textbf{Quantitative Comparison.} We also provide quantitative results in Table \ref{tab:comparison_table} as a complementary to the visual results. From Table \ref{tab:comparison_table},  our T-VAE model achieves  higher PSNR and SSIM scores and lower LPIPS scores, demonstrating the superiority of our model against the other models. It is worth noting that both our T-VAE model and the C-LSTM model use LSTMs to describe the temporal information in a sequence. However, our T-VAE model can encode the distribution of the generated data, which is better at dealing with major changes in the temporal progression of bone. Furthermore, our proposed edge-aware loss effectively allows the model to focus on the valid pixels in the images, thus providing more accurate predictions.

\subsection{Ablation Study}
We conduct an ablation study to evaluate the effects of different components and loss functions. Specifically, we compare our model using the proposed \textit{Edge-Aware} loss with the model using a plain \textit{Mean-Square-Error} (MSE) loss. Note that our full model takes 2D slices as the input and uses 2D convolutions in the encoders and decoders to extract features. As an alternative, we can instead adopt a volume composed of 48 2D slices as the input, and use 3D convolutions in the encoders and decoders to project the volume. We denote this variant of our model as T-VAE(3D) and denote our default version (using 2D convolution) as T-VAE(2D). The quantitative results of different versions of our model are shown in Table \ref{tab:ablation_table}.

\begin{figure}[t]
    \centering
    \includegraphics[width=\textwidth]{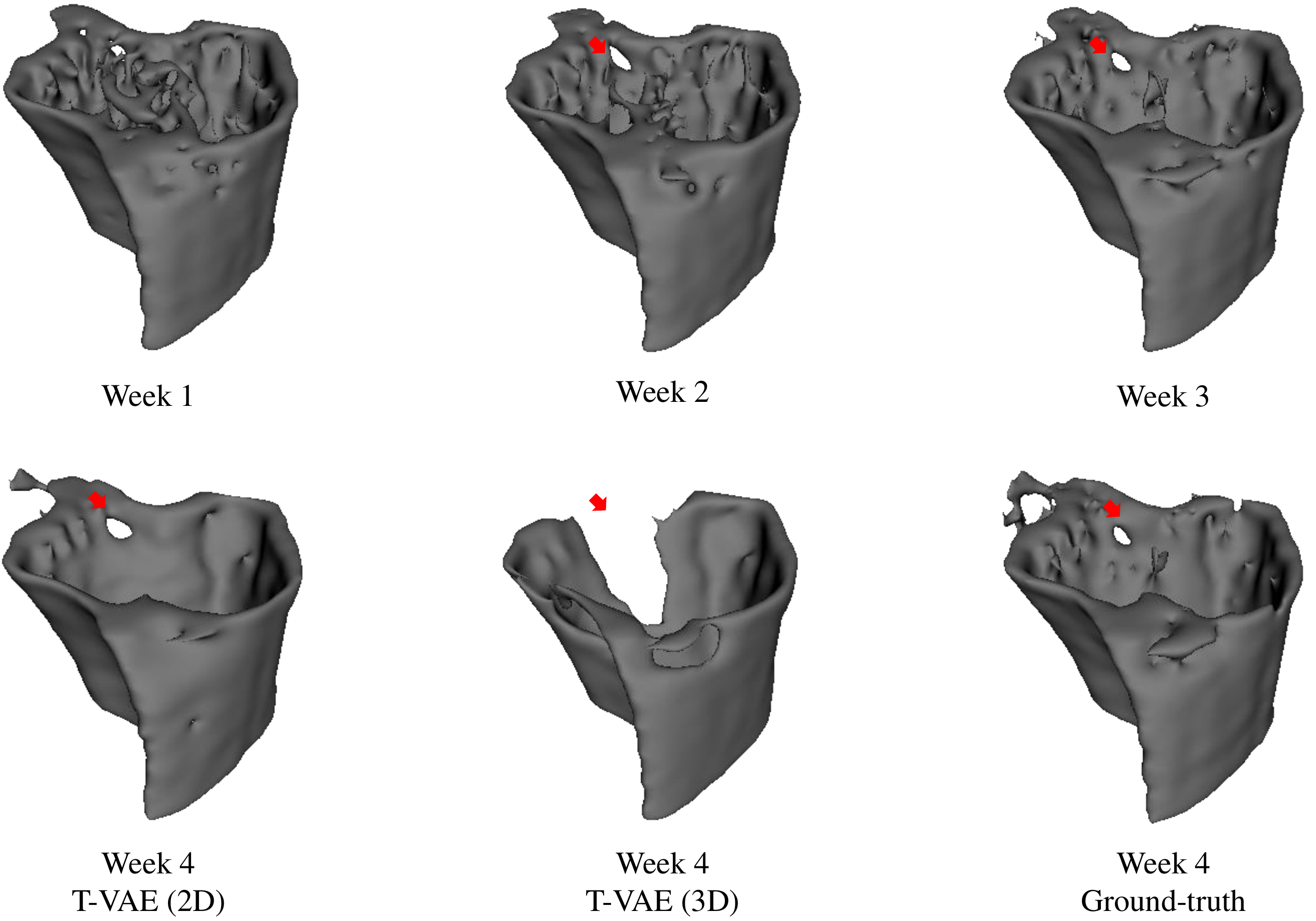}
    \caption{Visualization of the predicted volume of T-VAE(2D) and T-VAE(3D). T-VAE (2D) predicts the accurate locations and size of the lesion, while T-VAE (3D) over-predicts the lesion region, i.e., the lesion region is significantly larger than the lesion in the ground-truth. Red arrows indicate the lesion regions.}
    \label{fig:3d_viz}
\end{figure}

\input{tables/ablation}

\subsubsection{MSE Loss versus Edge-Aware Loss}
Concluded from Table \ref{tab:ablation_table}, utilizing the edge-aware loss during training, the T-VAE(2D) model achieves better scores than using the plain MSE loss. By factoring in the Gaussian mask of the first week's scan using Edge-aware loss, the T-VAE(2D) model is subsequently able to produce sharper and higher quality generations evidenced by the gains in the image reconstruction metrics PSNR and SSIM.

\subsubsection{T-VAE(2D) versus T-VAE(3D)}
We also compared our T-VAE(2D) version with T-VAE(3D) version. The only difference is that for the 2D version, we use 2D convolutions to extract features from each 2D image slice, while for the 3D version, we use 3D convolutions to extract features from each 3D volume (each volume of data contains 48 2D image slices). The encoder and decoder of the T-VAE(3D) model have similar architectures as the generator and discriminator of DCGAN \cite{dcgan}. Theoretically, the 3D model should be better at capturing the volumetric nature of the data than the 2D model. However, in practice, the T-VAE(2D) model outperforms the T-VAE(3D) model in terms of both quantitative performance shown in Table \ref{tab:ablation_table} and qualitative performance shown in Figure \ref{fig:3d_viz}. In Figure \ref{fig:3d_viz}, T-VAE(3D) over-predicts the lesion as indicated by the red arrows. In contrast, our T-VAE(2D) model predicts the lesion much more accurately. Moreover, volumes predicted by our T-VAE(2D) model contains more content details. These results indicate a sign of overfitting of the T-VAE(3D) model. We suspect the reason could be that the dataset does not have enough 3D volumes as training data in comparison to the training set of 2D slices, even when data augmentation is used. We plan to obtain more 3D volumes to address this issue.

To make the ablation study for the model architecture more solid, we include the result of another 3D version of our model. Two recent works \cite{GP-GAN,chen2020qsmgan} have used powerful 3D-GAN models for 3D data generation and achieved promising performance. Both of them have used a 3D U-Net as the generator. We hypothesize that a more effective architecture could compensate for the lack of data. The 3D U-Net is such a design that uses full skip-connections (``full'' means each layer in the encoder is connected to the corresponding layer in the decoder) to transfer knowledge from the encoder to the decoder, leading to a more effective usage of data. Therefore, we follow their architecture and use 3D U-Net 3D modeling instead of the plain 3D encoder-decoder architecture (the 3D DCGAN architecture). The results are shown in Table \ref{tab:ablation_table}. Using a 3D U-Net instead of our DCGAN architecture can improve the PSNR and LPIPS scores of our T-VAE 3D version. The reason could be that more content features are transferred from the previous frames to the future frames.

\subsection{Diversity of Prediction}
\begin{figure}[t]
    \centering
    \includegraphics[width=\columnwidth]{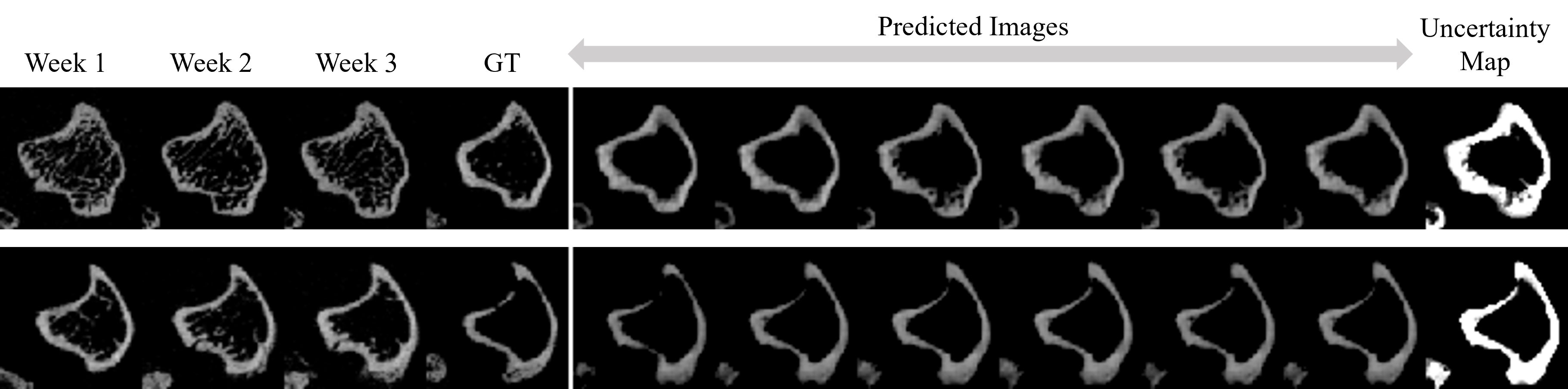}
    \caption{In each row, we show the slices of the first three weeks, the ground-truth slice in the fourth week, six predicted slices by sampling various $z_t$ of our model, and the uncertainty map of the predicted slices. We observe that the primary part of the image, i.e., the boundary of the bone (the tibial cortical bone) remains unchanged, while the other parts (the bone marrow regions) are diverse. Please zoom in to see the details of the predictions and the uncertainty map.}
    \label{fig:z_t}
\end{figure}

\begin{figure}[t]
    \centering
    \includegraphics[width=0.75\columnwidth]{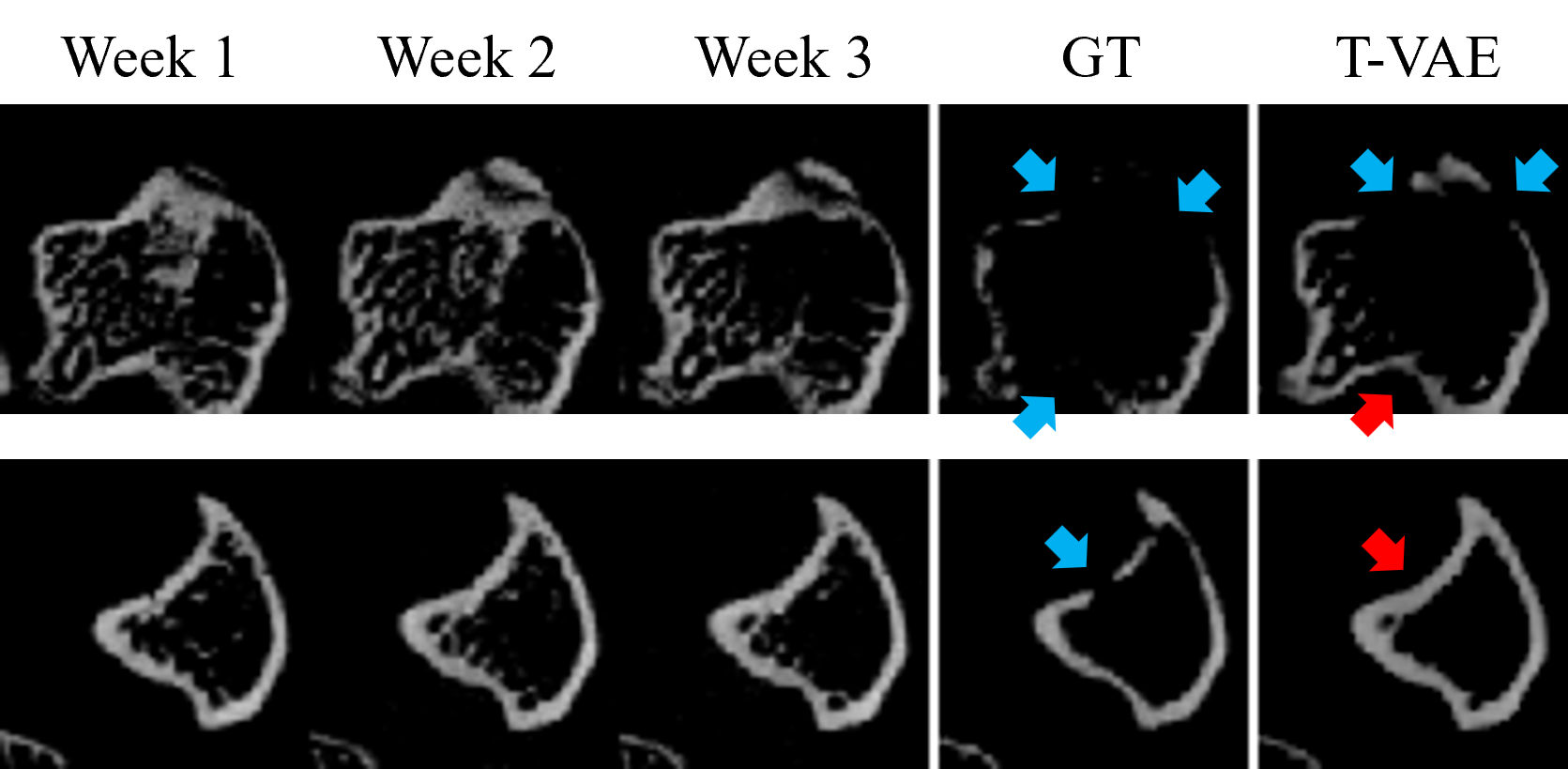}
    \caption{Failure samples generated by our method. In each row, we show the slices of the first three weeks, the ground-truth slice in week 4, and our predicted slice. Red arrows indicate wrong lesion predictions. Blue arrows indicate lesions in the ground-truth image or correct lesion predictions.}
    \label{fig:failure}
\end{figure}

Our model contains a stochastic component that samples the latent vector $z_t$ from an encoded latent distribution. Such a design is based on the stochastic nature of the progression of bone lesion. Given the bone slices of the previous weeks, there is still reasonable uncertainty on the bone state of the next week. The distribution intuitively reflects the range of plausible outcomes of the bone progression based off the previous weeks' scans $x_1, x_2, x_3$. Depending on the choice of $z_t$, the constitution of the resulting prediction $x'_{4}$ can vary as shown in Figure \ref{fig:z_t}. Interestingly, given different $z_t$, the boundary regions (the tibial cortical bone) of our predicted bone remain largely unchanged. Such regions are the key factors for a clinician to determine if there is lesion in the bone or not.  In the meanwhile, different contents are generated in the non-boundary areas (the bone marrow regions), showing the diversity of our results.

Figure \ref{fig:z_t} also shows the uncertainty map of the diverse predictions. The intensity of each pixel in the uncertainty map indicates the probability of the predicted pixel in that location to be non-zero. From the uncertainty map we observe that pixels in the boundary regions show a very high probability, while pixels in the non-boundary regions show a lower probability, i.e., our model can capture the core temporal patterns and generate pixels with a very high confidence and low uncertainty in the important regions (boundary regions). Our model also enables the diversity by generating pixels with a relatively lower confidence (higher uncertainty) in the non-boundary regions. 

\subsection{Failure Cases}
Figure \ref{fig:failure} shows two cases where our model does not make very accurate predictions. In the first row, our model is able to predict two of the lesions but fails to predict all the lesions. In the second row, our model fails to predict the lesion. In summary, our model sometimes cannot predict the lesion in a very accurate way when the lesion occurs in the fourth week but there is no lesion in the first three weeks. We will address these challenging cases in our future work.

%% file: tables/compare.tex
\begin{table*}[t]
\caption{Average PSNR, SSIM, and LPIPS score of predictions from different models. $\uparrow$ means higher is better. $\downarrow$ means lower is better. }

\small
\centering 
\begin{tabular}{ l c c c c}
\hline
Model & PSNR $\uparrow$ & SSIM $\uparrow$ & LPIPS $\downarrow$\\
\hline

C-LSTM &  22.81 & 0.786 & 0.087\\
3D-GAN & 21.98 & 0.760 & 0.095\\
PredRNN & 22.52 & 0.798 & 0.089 \\
T-VAE (2D) & \textbf{23.44} & \textbf{0.816} & \textbf{0.079}\\

\hline
\end{tabular}
\label{tab:comparison_table}
\end{table*}

%% file: tables/ablation.tex
\begin{table*}[t!]
\caption{Quantitative results of ablation study. $\uparrow$ means higher is better. $\downarrow$ means lower is better.}

\small
\centering 
\begin{tabular}{ l c c c c}
\hline
Models & PSNR $\uparrow$ & SSIM $\uparrow$ & LPIPS $\downarrow$ \\
\hline
T-VAE (2D) + MSE Loss & 22.57 & 0.785 & 0.083\\
T-VAE (2D) + \textit{Edge-Aware} Loss & {23.44} & {0.816} & {0.079}\\
T-VAE (3D-DCGAN) + MSE loss & 22.21 & 0.790 & 0.084 \\
T-VAE (3D-DCGAN) + \textit{Edge-Aware} Loss & 22.77 & 0.782 & 0.087 \\
T-VAE (3D-UNet) + \textit{Edge-Aware} Loss & 22.87 & 0.774 & 0.085 \\
\hline
\end{tabular}
\label{tab:ablation_table}
\end{table*}

%% file: discussion.tex
\section{Discussion}

In this study, we showed that bone lesions can be reasonably predicted and visualized by utilizing deep learning models. Given the scans from the previous three time points, our T-VAE model generated diverse and plausible images for the fourth time point. Our model outperformed various future prediction models including the 3D-GAN, the Convolutional LSTM and the PredRNN, as measured in various reconstruction similarity metrics such as PSNR and SSIM, as well as perceptual metrics such as LPIPS.

During the prediction process of the model, a latent vector, $z_t$, is sampled from a learned distribution. The sampling of this vector is stochastic. The choice of $z_t$ affects the appearance of the final prediction. On a high level, the latent vector is supposed to represent the stochastic changes within the bone cortex between the discrete CT images of each week. Although the diversity of the latent vector can result in a distribution of plausible outcomes of plausible predictions as evidenced by Figure \ref{fig:z_t}, the most important regions such as the boundary regions that can indicate the lesions usually remain unchanged.

Our study explored the potential of using deep generative models to predict the occurrence and progression of bone lesions with a reasonable accuracy using sequential scans in a preclinical CT dataset. Future studies should be performed using datasets containing images from different breast cancer subtypes or other cancer types. The breast cancer used in the current dataset was the aggressive triple-negative breast cancer. The approach will need to be tested on clinical sequential datasets, which are expected to present several technical challenges. Due to the health risk associated with radiation, the scan frequency and subsequently the intervals between scans may vary from patients to patients. How to model temporal patterns with varying time intervals remains a challenge. 

Although the resolution of clinical CT scans is typically lower than that of preclinical CT scans, the thicker human bone cortex could compensate this limitation during the detection of bone lesions and their effects on bone structural compromise. 

Despite these challenges, the present study provides the foundation of a deep learning framework, which could lead to early detection of osteolytic bone lesions and the ability of predict fracture risk associated with metastatic breast cancer. The approach can be applied to other cancer bone metastasis. Our long-term goal is to assist planning and evaluating treatment strategies to prevent painful or even fatal adverse skeletal related events in cancer patients.

%% file: method.tex
\section{Materials and Method}
\subsection{Data Acquisition and Preprocessing} 
The dataset consists of CT image scans of the metaphysis of tibiae of adult female mice, which received either breast cancer cells or phosphate-buffered saline (PBS), followed by weekly scans using a CT scanner with a voxel size of 7 micron.
Each bone volume is a data point $x \in \mathbb{R}^{100\times100\times48}$ with height and width $100$ and number of slices $48$. 
The data were processed using SITK \cite{sitk} to register the CT scan. Each resulting 2D slice is a gray-scale, single-channel image. The slice was finally  resized to be  $64\times64\times1$ and then normalized. There were $251$ mice and approximately $100$ of them had no tumor. The ratio of slices with lesion emergence and lack of lesion emergence in non-tumor mice was explicitly adjusted in the experiments. Each mouse had 3-6 weekly 3D scans. In the current study, we only considered samples with 4 weeks of data points to ensure temporal consistency.

Our preprocessing protocols are as following. We utilize an 80:20 split of the samples containing only 4 weeks of data points, resulting in 88 CT volume sequences (each sequence contains 4 volume data points) in the training set and 23 CT volume sequences in the test set. Basic data augmentations are applied to the training set including random horizontal flips on the training set. For more detail on the training parameters, refer to subsection \ref{training}.

It is worth noting that there is currently no other publicly available datasets that contain CT scans of bone with bone lesions and have temporal sequences of CT scans rather than single discrete scans without temporal progression. Therefore, we do not benchmark our method on other datasets.


\begin{figure}
    \centering
    \includegraphics[width=\columnwidth]{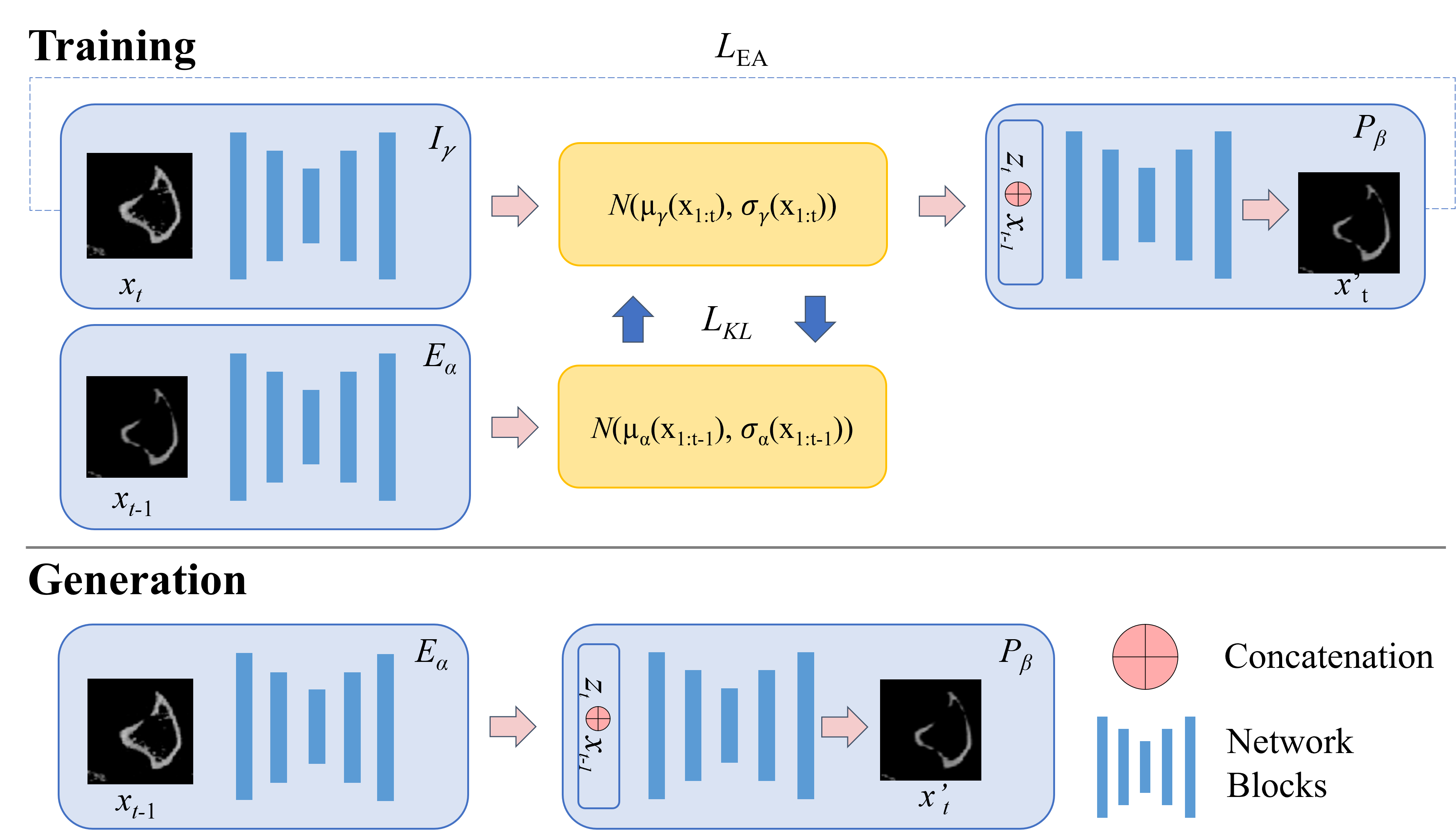}
    \caption{The training and generation process of T-VAE.}
    \label{fig:framework}
\end{figure}

\subsection{Method}

\subsubsection{Overview of Temporal Variational Auto-Encoders (T-VAE)}
Our model is derived from the previous work \cite{svg-lp}. Given data at the previous time steps $x_{1},..x_{t-1}$, our T-VAE model predicts data at the $t$-th step $x_{t}$. It is composed of three modules: a prior estimator $E_{\alpha}$, a future predictor $P_{\beta}$, and an latent inference model $I_{\gamma}$, where $\alpha$, $\beta$ and $\gamma$ are learnable parameters of these models, respectively. The prior estimator learns a prior representation that describe the distribution of the data in the current time step. It encodes the uncertainty of the temporal progression of data sequences. The predictor takes the prior and data from the previous time steps $x_{1},..x_{t-1}$ to generate data in the current time step $x_t$. To facilitate the training, the inference model is used to map the sequence of data to latent representations. An illustration of our framework is shown in Figure \ref{fig:framework}. We describe the modules in detail as below.

\subsubsection{Prior Estimator}
The vanilla variational auto-encoder adopts the standard Gaussian $\mathcal{N}(0, 1)$ distribution as the prior to model the variance in the data. For temporal data, however, this is not the optimal encoding scheme. One primary reason is that the uncertainty at the current step is not fully stochastic, but influenced by the states of the history data. Therefore, we introduce the prior estimator model $E_{\alpha}$. It is a recurrent model that calculates the prior of each time step $z_{\alpha}(t)$ as a Gaussian distribution conditioned on data at the previous steps. At each time step $t$, we have
\begin{equation}
    \mathbf{z}_{\alpha}(t)  \sim \mathcal{N}\left(\mu_{\alpha}(x_{1:t-1}), \sigma_{\alpha}({x_{1:t-1}})\right)
\end{equation}
In the following, we use $\mu_{\alpha}(t)$ and $\sigma_{\alpha}(t)$ to represent $\mu_{\alpha}(x_{1:t-1})$ and $\sigma_{\alpha}({x_{1:t-1}})$ for simplicity, respectively. $\mu_{\alpha}(t)$ and $\sigma_{\alpha}(t)$ are calculated using our prior estimator $E_{\alpha}$. We have

\begin{equation}
    \mu_{\alpha}(t), \sigma_{\alpha}(t) = E_{\alpha}(x_{t-1}, h_{\alpha}(t-1))
\end{equation}
where $h_{\alpha}(t-1)$ is the hidden state in the prior estimator at the $t-1$ step.

\subsubsection{Future Predictor}
Our future predictor $P_{\beta}$ is a recurrent model based on LSTM networks and the variational auto-encoder (VAE). Specifically, it is composed of an encoder to extract features from each frame, an LSTM model to capture the temporal information in a sequence and a decoder that maps the features back to image frames. 

At each time step $t$, our predictor $P$ takes the hidden state at the previous time step $h_{\beta}(t-1)$, and the latent representation at the current time step $\mathbf{z}_{\gamma}(t)$, to generate the current frame ${x'}_t$. The hidden state $h_{\beta}(t-1)$ encodes the temporal pattern among the data at the previous time steps $x_{1},..x_{t-1}$. During training, the latent representation $\mathbf{z}_{\gamma}(t)$ is computed by the latent inference model $I_{\gamma}$, which we will formulate later. We have

\begin{equation}\label{eq:predictor}
    {x'}_t = P_{\beta}(h_{\beta}(t-1), \mathbf{z}_{\gamma}(t))
\end{equation}

\subsubsection{Latent Inference Model}
Our latent inference model $I_{\gamma}$ is only used during training. It is composed of an encoder that extracts features of each frame and an LSTM model that models the temporal patterns within a sequence. The latent inference model is used to calculate the posterior $p(z_t|x_{1:t})$ used in the future predictor. We have

\begin{equation}
    \mathbf{z}_{\gamma}(t) = I_{\gamma}(h_{\gamma}(t-1), x_{t})
\end{equation}
where $h_{\gamma}(t-1)$ is the hidden state of the inference model at time step $t$, representing the history temporal information of the data sequence $x_{1},..x_{t-1}$. $x_t$ is the ground-truth frame at the $t$-th time step. Similar to the prior in the prior estimator, the posterior representation $\mathbf{z}_{\gamma}(t)$ also follows a conditional Gaussian distribution. We have

\begin{equation}
    \mathbf{z}_{\gamma}(t)  \sim \mathcal{N}\left(\mu_{\gamma}(x_{1:t}), \sigma_{\gamma}({x_{1:t}})\right)
\end{equation}
where $\mu_{\gamma}(x_{1:t})$ and $\sigma_{\gamma}({x_{1:t}})$ are the outputs of the LSTM in our latent inference model at time step $t$.

\subsubsection{Learning Objectives}
During training, We learn the modules by minimizing the reconstruction loss (implemented with mean-square-error) between the predicted frame and the ground-truth frame at time step $t$. We have:

\begin{equation}\label{eq:recon}
    \mathcal{L}_{recon} = \sum_{t = 1}^{T}(x'_t - x_t)^{2}
\end{equation}
where $T$ is the total number of time steps in each training sequence. Note that using only the $\mathcal{L}_{recon}$ will result in a deterministic model that cannot provide any stochasticity. The inference model may also degrade to memorize the ground-truth data $x_t$ at the $t$-th step. To address these issues, we follow the idea of the variational auto-encoder, and enforce the posterior representation of the inference model $\mathbf{z}_{\gamma}(t)$ to be close to the prior representation of the prior estimator $\mathbf{z}_{\alpha}(t)$ at each time step $t$. We use the KL-divergence loss to achieve our goal. We have

\begin{equation}
    \mathcal{L}_{KL} = \sum_{t = 1}^{T}D_{\mathrm{KL}}(\mathbf{z}_{\gamma}(t) \| \mathbf{z}_{\alpha}(t))
\end{equation}
where $D_{\mathrm{KL}}(p \| q)$ denotes the Kullback-Leibler divergence (KL-Divergence), formulated as
\begin{equation}
    D_{\mathrm{KL}}(p \| q)=\sum^{N}_{i} p_{i} \log _{2} \frac{p_{i}}{q_{i}}
\end{equation}
where $p$ and $q$ both denote probability distribution and $q$ denotes a ``target'' probability distribution \cite{shlens2014notes}.

At each time step, the KL-divergence loss forces the inference model to match prior distributions rather than memorizing history data, so that the predictor that is conditioned on the posterior representation can learn new patterns which does not exists in the previous data. 

\begin{figure*}
    \centering
    \includegraphics[width=0.7\textwidth]{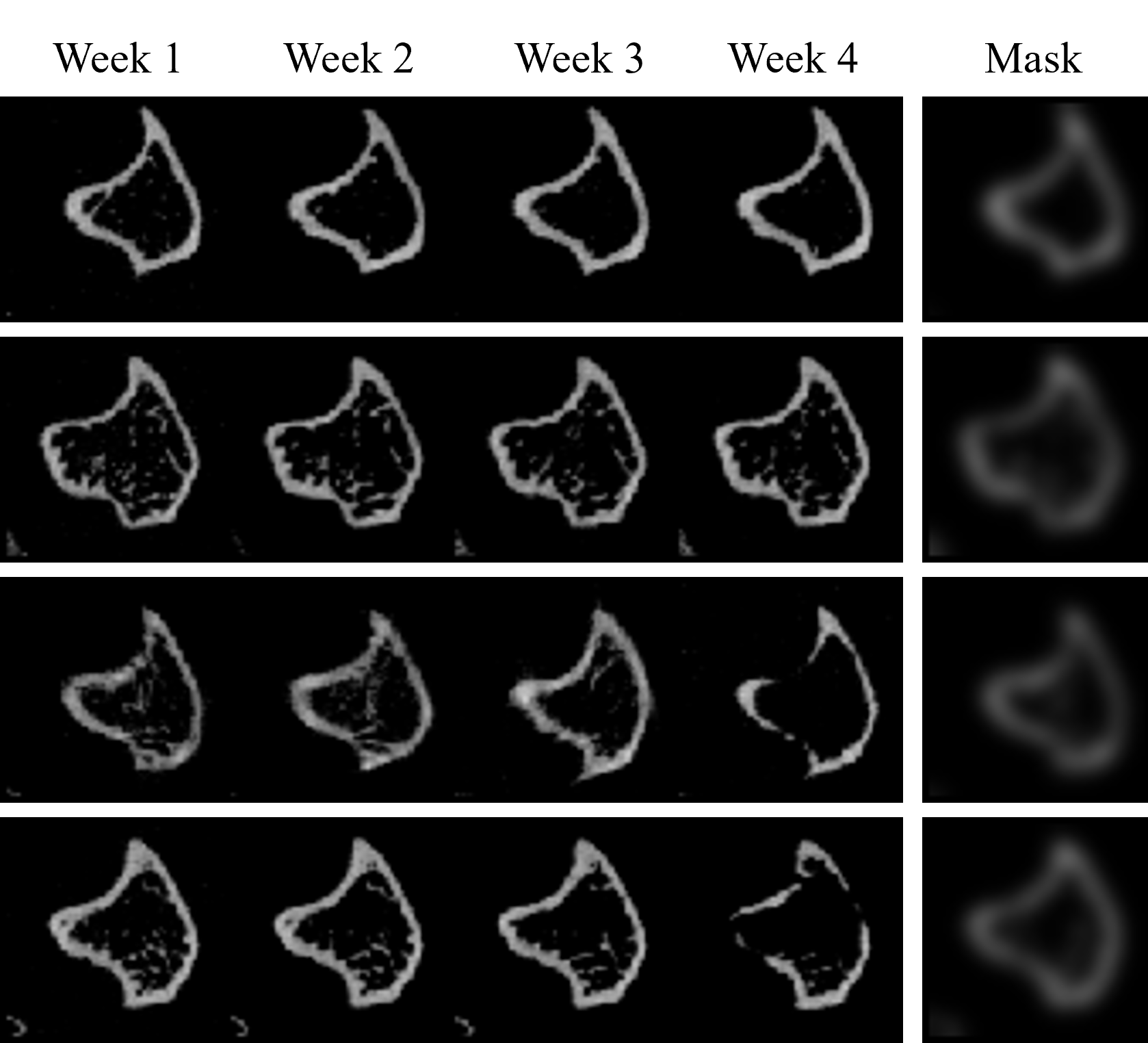}
    \caption{Visualization of the Gaussian masks $G$ that are used in the edge-aware loss.}
    \label{fig:masks}
\end{figure*}

\subsubsection{Edge-Aware Learning}

Since the pixels of a CT imaging data usually exhibit sparseness, i.e., after normalization, most of the pixels are close to 0 and only a small portion of the pixels are valid. Directly adopting the reconstruction loss in equation \ref{eq:recon} can lead to inaccurate prediction or blurry results. To address this issue, we encourage the model to pay more attention to the valid pixels in the CT slices/frames. These pixels form the edges and boundary of the bone structure. A typical solution is to use the focal loss \cite{focal_loss}. However, the conventional focal loss is usually used with the cross-entropy loss or other losses with similar forms. It is not proper enough to use it on the reconstruction loss. We propose an edge-aware loss, borrowing the idea from focal loss, but with a rather simple yet effective form. Notice that in our task the significant edges or boundary are usually already provided by data at the first week, as in the first week, the cancer is just injected to the bones, and it takes time for the bone lesion induced by cancer. Therefore, data in the first week usually preserves all the edges and boundary. It is indeed a good mask that can guide the model where to pay attention to. One remaining issue is that data in the first week is too sharp to serve as a mask, which can harm the optimization of the modules. To solve this issue, we blur the data using Gaussian kernels and then use it as the mask indicating important regions. We modify the reconstruction loss in equation \ref{eq:recon} as follows:
\begin{equation}
    \mathcal{L}_{EA} = \sum_{t = 1}^{T}(1 + \lambda G(x_{1}))(x'_t - x_t)
\end{equation}
where $G(x_{1})$ denotes to the Gaussian blurred edge mask, $\lambda$ that ranges from 0 to 1 is the weight for the edge mask and $\sigma$ is the standard deviation for the Gaussian kernel. Figure \ref{fig:masks} shows examples of our generated masks.

The total loss is a combination of the edge-aware reconstruction loss and the KL-divergence loss. We have
\begin{equation}
    \mathcal{L} = \mathcal{L}_{EA} + 0.0001 * \mathcal{L}_{KL}.
\end{equation}

\subsubsection{Testing Phase}
Notably, the models adopted in the testing phase differ from those adopted in the training phase. The primary difference is that the latent inference model cannot be used, as it uses ground-truth data $x_t$ to calculate the posterior latent representation $\mathbf{z}_{\gamma}(t)$. In the future predictor, instead of using the posterior representation to calculate the final predicted frame at time step $t$, we sample latent representations from the prior learned by the prior estimator. We modify equation \ref{eq:predictor} to the following form:

\begin{equation}\label{eq:predictor_modified}
    {x'}_t = P_{\beta}(h_{\beta}(t-1), \mathbf{z}_{\alpha}(t))
\end{equation}

In this way, we can generate a new frame at time step $t$ using data at the previous time steps $x_{1},..x_{t-1}$ and the sampled latents from the learned prior. 

\subsubsection{Architecture and Training Parameters} \label{training}
Our T-VAE model follows the encoder-decoder architecture of the vanilla VAE. The Encoder and Decoders of T-VAE use a DCGAN \cite{dcgan} discriminator and generator architecture, respectively \cite{dcgan}. To enrich the content information in the output slices, we connect a few layers in the encoder to the corresponding layers in the decoder (full skip-connections for each layer are not applied). The default version of our model uses 2D convolutions in the encoders and decoders. The input and output of the model at each time step is a single 2D slice. The decoder has a sigmoid output layer. The dimension of the latent vectors is $|z_t| = 10$ and the dimension of the output vector of the encoder is given by $g_{dim} = 128$. The model is inspired by the stochastic video generations (SVG) model \cite{svg-lp} . We train the T-VAE model with a batch size of $48$, for a max of $200$ epochs and early stopping. We use similar training parameters for 3D-GAN, C-LSTM and PredRNN.  During training, we apply a random crop of $96 \times 96$, as well and then resize the slice to $64 \times 64$. During testing, we input to the model 2D slices of the first three weeks (all the three slices are from the same location of the given 3D volume), then our model generates the slice in the fourth week of that location. The resulting slices form a predicted 3D volume. 

As a variant of our default model, our model can take a sequence of 3D volumes instead of 2D slices as the inputs. We denote this variant as T-VAE(3D). The inputs to this model are the first three weeks of 3D volumes (each composed of 48 2D slices). We replace the 2D convolutions of the T-VAE model with 3D convolutions in the encoders and decoders. The other parts of the model remain unchanged. The output of our model is the predicted 3D volume containing 48 2D slices. We use a smaller batch size of 4 but keep the same number of max epochs.

 We use the Adam optimizer with momentum $\beta_1 = 0.9$ and a learning rate of $0.002$. We conduct a hyper-parameter search on $\lambda$ and $\sigma$ values of the Gaussian mask $G$ and find that $\lambda=1.0$ and $\sigma=5.0$ yields the best results. The experiments are implemented with PyTorch and run on a NVIDIA\textsuperscript{\textcopyright} 1080Ti GPU.

%% file: animalstudies.tex
\subsection{Animal and Human Studies}
Animal experiments were approved by the Institutional Animal Care and Use Committee (IACUC) and conducted in an accredited animal facility at the University of Delaware. In brief, female C57BL/6J mice (Jackson Laboratory, Bar Harbor, ME, USA) were inoculated with \textit{Mus musculus} mesenchymal-like Py8119 breast cancer cells (ATCC, Manassas, VA , USA, CRL-3278\textsuperscript{TM}) through intratibial injection at the age of 14 weeks old. Separate animals injected with PBS served as non-tumor control group. Tumor-inoculated animals started developing osteolytic bone lesions 2 weeks after injection. Animals were also subjected to physical activities such as local cyclic compression of the tibia and treadmill running, which regulated tumor growth and the progression of bone lesions. The details of the experimental studies have been published \cite{wang2021}. \textit{In vivo} micro CT ($\mu$CT) scans were conducted weekly for 3-6 weeks using SKYSCAN\textsuperscript{\textregistered} 1276 (Bruker, Kontich, Belgium). The imaging settings included 900 ms exposure time, 200 mA current and 50kVp, a 0.5mm Al filter and a 7$\mu$m voxel size. Images were reconstructed using NRecon\textsuperscript{\textregistered} software (Bruker). For all the images, the volume of interest was the proximal tibial metaphysis of 2.1 mm thick (300 slices) below the growth plate. To ensure consistency of bone orientation, the weekly scans for each bone were registered using its first scan, which was registered using a common reference tibia for all the animals. 

%% file: acknowledgements.tex
\section*{Authors' Contributions}
W. Xiong and N. Yeung designed and implemented the deep learning framework, and conducted the experiments. W. Xiong, H. Liao and J. Luo conceived the idea of the edge-aware loss and other implementation ideas. S. Wang imaged and collected the original data set, and helped with visualization of the 3D volumes. L. Wang conceived and designed the animal study, from which the data set was obtained. J. Luo and L. Wang conceived this study, and provided guidance and feedback. All authors contributed to the writing of the manuscript.

\section*{Conflicts of Interest}
The authors have declared no conflicts of interest.

\section*{Acknowledgements}
The animal work and micorCT scanning was partially supported by the National Institutes of Health (R01AR054385 to L. Wang). The image prediction work was partially supported by the National Science Foundation (1704337 to J. Luo).

\section*{Data Availability}
The micro-CT data are available on a data server maintained at the University of Delaware. Free download is available upon request.  

%% file: BME Frontiers Final/main.bib
@Article{metastasis,
   Author="Mundy, G. R. ",
   Title="{{M}etastasis to bone: causes, consequences and therapeutic opportunities}",
   Journal="Nat Rev Cancer",
   Year="2002",
   Volume="2",
   Number="8",
   Pages="584--593",
   Month="Aug"
}

@article{focal_loss,
  author    = {Tsung{-}Yi Lin and
               Priya Goyal and
               Ross B. Girshick and
               Kaiming He and
               Piotr Doll{\'{a}}r},
  title     = {Focal Loss for Dense Object Detection},
  journal   = {CoRR},
  volume    = {abs/1708.02002},
  year      = {2017},
  url       = {http://arxiv.org/abs/1708.02002},
  archivePrefix = {arXiv},
  eprint    = {1708.02002},
  bibsource = {dblp computer science bibliography, https://dblp.org}
}

@Article{bone_events,
   Author="Clemons, M.  and Gelmon, K. A.  and Pritchard, K. I.  and Paterson, A. H. ",
   Title="{{B}one-targeted agents and skeletal-related events in breast cancer patients with bone metastases: the state of the art}",
   Journal="Curr Oncol",
   Year="2012",
   Volume="19",
   Number="5",
   Pages="259--268",
   Month="Oct"
}

@article{pet_ct_mri_prediction,
  title={Prediction of early metastatic disease in experimental breast cancer bone metastasis by combining PET/CT and MRI parameters to a Model-Averaged Neural Network},
  author={Ellmann, Stephan and Seyler, Lisa and Evers, Jochen and Heinen, Henrik and Bozec, Aline and Prante, Olaf and Kuwert, Torsten and Uder, Michael and B{\"a}uerle, Tobias},
  journal={Bone},
  volume={120},
  pages={254--261},
  year={2019},
  publisher={Elsevier}
}

@article{murinemodel,
  title={Murine models of breast cancer bone metastasis},
  author={Wright, Laura E and Ottewell, Penelope D and Rucci, Nadia and Peyruchaud, Olivier and Pagnotti, Gabriel M and Chiechi, Antonella and Buijs, Jeroen T and Sterling, Julie A},
  journal={BoneKEy reports},
  volume={5},
  year={2016},
  publisher={Nature Publishing Group}
}

@article{imaging_dia,
  title={Imaging diagnosis of metastatic breast cancer},
  author={Pesapane, Filippo and Downey, Kate and Rotili, Anna and Cassano, Enrico and Koh, Dow-Mu},
  journal={Insights into imaging},
  volume={11},
  number={1},
  pages={1--14},
  year={2020},
  publisher={Springer}
}

@Article{cancer_stat,
   Author="Siegel, R. L.  and Miller, K. D.  and Jemal, A. ",
   Title="{{C}ancer statistics, 2020}",
   Journal="CA Cancer J Clin",
   Year="2020",
   Volume="70",
   Number="1",
   Pages="7--30",
   Month="01"
}

@inproceedings{ronneberger2015u,
  title={U-net: Convolutional networks for biomedical image segmentation},
  author={Ronneberger, Olaf and Fischer, Philipp and Brox, Thomas},
  booktitle={International Conference on Medical image computing and computer-assisted intervention},
  pages={234--241},
  year={2015},
  organization={Springer}
}

@inproceedings{li2019patch,
  title={Patch transformer for multi-tagging whole slide histopathology images},
  author={Li, Weijian and Nguyen, Viet-Duy and Liao, Haofu and Wilder, Matt and Cheng, Ke and Luo, Jiebo},
  booktitle={International Conference on Medical Image Computing and Computer-Assisted Intervention},
  pages={532--540},
  year={2019},
  organization={Springer}
}

@inproceedings{li2020structured,
  title={Structured landmark detection via topology-adapting deep graph learning},
  author={Li, Weijian and Lu, Yuhang and Zheng, Kang and Liao, Haofu and Lin, Chihung and Luo, Jiebo and Cheng, Chi-Tung and Xiao, Jing and Lu, Le and Kuo, Chang-Fu and others},
  booktitle={European Conference on Computer Vision},
  pages={266--283},
  year={2020},
  organization={Springer}
}

@misc{wang2021predrnn,
      title={{PredRNN}: A Recurrent Neural Network for Spatiotemporal Predictive Learning}, 
      author={Wang, Yunbo and Wu, Haixu and Zhang, Jianjin and Gao, Zhifeng and Wang, Jianmin and Yu, Philip S and Long, Mingsheng},
      year={2021},
      eprint={2103.09504},
      archivePrefix={arXiv},
}

@inproceedings{zhang2018unreasonable,
  title={The unreasonable effectiveness of deep features as a perceptual metric},
  author={Zhang, Richard and Isola, Phillip and Efros, Alexei A and Shechtman, Eli and Wang, Oliver},
  booktitle={Proceedings of the IEEE conference on computer vision and pattern recognition},
  pages={586--595},
  year={2018}
}

@misc{dcgan,
      title={Unsupervised Representation Learning with Deep Convolutional Generative Adversarial Networks}, 
      author={Alec Radford and Luke Metz and Soumith Chintala},
      year={2016},
      eprint={1511.06434},
      archivePrefix={arXiv},
      primaryClass={cs.LG}
}

@misc{sitk, title={SimpleITK Image-Analysis Notebooks: a Collaborative Environment for Education and Reproducible Research}, 
 journal={Journal of digital imaging}, publisher={U.S. National Library of Medicine}, author={R;, Yaniv Z;Lowekamp BC;Johnson HJ;Beare}, year={2019}, month={Dec}}

@misc{zhang2019spatiotemporal,
      title={Spatio-Temporal Convolutional LSTMs for Tumor Growth Prediction by Learning 4D Longitudinal Patient Data}, 
      author={Ling Zhang and Le Lu and Xiaosong Wang and Robert M. Zhu and Mohammadhadi Bagheri and Ronald M. Summers and Jianhua Yao},
      year={2019},
      eprint={1902.08716},
      archivePrefix={arXiv},
      primaryClass={cs.CV}
}

@misc{shlens2014notes,
      title={Notes on Kullback-Leibler Divergence and Likelihood}, 
      author={Jonathon Shlens},
      year={2014},
      eprint={1404.2000},
      archivePrefix={arXiv},
      primaryClass={cs.IT}
}

@article{kingma2013auto,
  title={Auto-encoding variational bayes},
  author={Kingma, Diederik P and Welling, Max},
  journal={arXiv preprint arXiv:1312.6114},
  year={2013}
}

@article{GP-GAN,
title = "GP-GAN: Brain tumor growth prediction using stacked 3D generative adversarial networks from longitudinal MR Images",
journal = "Neural Networks",
volume = "132",
pages = "321 - 332",
year = "2020",
issn = "0893-6080",
doi = "https://doi.org/10.1016/j.neunet.2020.09.004",
url = "http://www.sciencedirect.com/science/article/pii/S0893608020303270",
author = "Ahmed Elazab and Changmiao Wang and Syed Jamal Safdar Gardezi and Hongmin Bai and Qingmao Hu and Tianfu Wang and Chunqi Chang and Baiying Lei",
}

@article{chen2020qsmgan,
  title={QSMGAN: improved quantitative susceptibility mapping using 3D generative adversarial networks with increased receptive field},
  author={Chen, Yicheng and Jakary, Angela and Avadiappan, Sivakami and Hess, Christopher P and Lupo, Janine M},
  journal={NeuroImage},
  volume={207},
  pages={116389},
  year={2020},
  publisher={Elsevier}
}

@article{vae,
   title={An Introduction to Variational Autoencoders},
   volume={12},
   ISSN={1935-8245},
   url={http://dx.doi.org/10.1561/2200000056},
   DOI={10.1561/2200000056},
   number={4},
   journal={Foundations and Trends® in Machine Learning},
   publisher={Now Publishers},
   author={Kingma, Diederik and Welling, Max},
   year={2019},
   pages={307–392}
}

@article{goodfellow2014generative,
  title={Generative adversarial nets},
  author={Goodfellow, Ian and Pouget-Abadie, Jean and Mirza, Mehdi and Xu, Bing and Warde-Farley, David and Ozair, Sherjil and Courville, Aaron and Bengio, Yoshua},
  journal={Advances in neural information processing systems},
  volume={27},
  year={2014}
}

@inproceedings{xiong2020fine,
  title={Fine-grained image-to-image transformation towards visual recognition},
  author={Xiong, Wei and He, Yutong and Zhang, Yixuan and Luo, Wenhan and Ma, Lin and Luo, Jiebo},
  booktitle={Proceedings of the IEEE/CVF Conference on Computer Vision and Pattern Recognition},
  pages={5840--5849},
  year={2020}
}

@article{svg-lp,
  author    = {Emily Denton and
               Rob Fergus},
  title     = {Stochastic Video Generation with a Learned Prior},
  journal   = {CoRR},
  volume    = {abs/1802.07687},
  year      = {2018},
  url       = {http://arxiv.org/abs/1802.07687},
  archivePrefix = {arXiv},
  eprint    = {1802.07687},
  bibsource = {dblp computer science bibliography, https://dblp.org}
}

@article{wang2021,
  title={Moderate tibial loading and treadmill running, but not overloading, protect adult murine bone from destruction by metastasized breast cancer},
  author={Wang, Shubo and Pei, Shaopeng and Wasi, Murtaza and Parajuli, Ashutosh and Yee, Albert and You, Lidan and Wang, Liyun},
  journal={Bone},
  pages={116100},
  year={2021},
  publisher={Elsevier}
}

@inproceedings{xiong2018learning,
  title={Learning to generate time-lapse videos using multi-stage dynamic generative adversarial networks},
  author={Xiong, Wei and Luo, Wenhan and Ma, Lin and Liu, Wei and Luo, Jiebo},
  booktitle={Proceedings of the IEEE Conference on Computer Vision and Pattern Recognition},
  pages={2364--2373},
  year={2018}
}

@article{trans-axial-rigidity,
  title={Treatment planning and fracture prediction in patients with skeletal metastasis with CT-based rigidity analysis},
  author={Nazarian, Ara and Entezari, Vahid and Zurakowski, David and Calderon, Nathan and Hipp, John A and Villa-Camacho, Juan C and Lin, Patrick P and Cheung, Felix H and Aboulafia, Albert J and Turcotte, Robert and others},
  journal={Clinical Cancer Research},
  volume={21},
  number={11},
  pages={2514--2519},
  year={2015},
  publisher={AACR}
}

@article{FEM2018,
  title={Pathological fracture risk assessment in patients with femoral metastases using CT-based finite element methods. A retrospective clinical study},
  author={Sternheim, Amir and Giladi, Ornit and Gortzak, Yair and Drexler, Michael and Salai, Moshe and Trabelsi, Nir and Milgrom, Charles and Yosibash, Zohar},
  journal={Bone},
  volume={110},
  pages={215--220},
  year={2018},
  publisher={Elsevier}
}

@article{damron2020fracture,
  title={Fracture risk assessment and clinical decision making for patients with metastatic bone disease},
  author={Damron, Timothy A and Mann, Kenneth A},
  journal={Journal of Orthopaedic Research{\textregistered}},
  volume={38},
  number={6},
  pages={1175--1190},
  year={2020},
  publisher={Wiley Online Library}
}

@article{eggermont2018can,
  title={Can patient-specific finite element models better predict fractures in metastatic bone disease than experienced clinicians? Towards computational modelling in daily clinical practice},
  author={Eggermont, Florieke and Derikx, LC and Verdonschot, Nico and Van Der Geest, ICM and De Jong, MAA and Snyers, An and Van Der Linden, YM and Tanck, Esther},
  journal={Bone \& joint research},
  volume={7},
  number={6},
  pages={430--439},
  year={2018},
  publisher={The British Editorial Society of Bone and Joint Surgery London}
}

@article{FEM2012,
  title={The assessment of the risk of fracture in femora with metastatic lesions: comparing case-specific finite element analyses with predictions by clinical experts},
  author={Derikx, Loes C and van Aken, Jantien B and Janssen, Dennis and Snyers, An and van der Linden, Yvette M and Verdonschot, Nico and Tanck, Esther},
  journal={The Journal of bone and joint surgery. British volume},
  volume={94},
  number={8},
  pages={1135--1142},
  year={2012},
  publisher={The British Editorial Society of Bone and Joint Surgery}
}

@article{convlstm,
  author    = {Xingjian Shi and
               Zhourong Chen and
               Hao Wang and
               Dit{-}Yan Yeung and
               Wai{-}Kin Wong and
               Wang{-}chun Woo},
  title     = {Convolutional {LSTM} Network: {A} Machine Learning Approach for Precipitation
               Nowcasting},
  journal   = {CoRR},
  volume    = {abs/1506.04214},
  year      = {2015},
  url       = {http://arxiv.org/abs/1506.04214},
  eprinttype = {arXiv},
  eprint    = {1506.04214},
  bibsource = {dblp computer science bibliography, https://dblp.org}
}
